\documentclass[%
 aip,
 amsmath,amssymb,
 reprint,%
 nofootinbib,
]{revtex4-1}

\usepackage{graphicx}
\usepackage{dcolumn}
\usepackage{bm}
\usepackage{siunitx}
\usepackage[utf8]{inputenc}
\usepackage[T1]{fontenc}
\usepackage{mathptmx}
\usepackage{hyperref}
\usepackage{graphicx} 
\usepackage[caption=false]{subfig}
\hypersetup{
     colorlinks   = true,
     citecolor    = blue,
     linkcolor    = blue}
\usepackage{tcolorbox}
\begin{document}

\preprint{AIP/123-QED}

\title{Modeling of Particle Transport, Neutrals and Radiation in Magnetically-Confined Plasmas with Aurora}

\author{F. Sciortino}
\email{sciortino@psfc.mit.edu}
\thanks{pronouns: he/him/his}
\affiliation{ 
Massachusetts Institute of Technology, Cambridge, MA, 02139, USA
}%
\author{T. Odstr\v{c}il}
\affiliation{ 
General Atomics, P.O. Box 85608, San Diego, CA 92186-5608, USA
}%
\author{A. Cavallaro}
\affiliation{ 
Massachusetts Institute of Technology, Cambridge, MA, 02139, USA
}%

\author{S. Smith}
\author{O. Meneghini}
\affiliation{ 
General Atomics, P.O. Box 85608, San Diego, CA 92186-5608, USA
}%

\author{R. Reksoatmodjo}
\affiliation{ 
Department of Physics, College of William \& Mary, Williamsburg, VA, USA
}%

\author{O. Linder}
\affiliation{ 
Max-Planck-Institut für Plasmaphysik, Boltzmannstraße 2, D-85748 Garching, Germany
}%

\author{J. D. Lore}
\affiliation{ 
Oak Ridge National Laboratory, Oak Ridge, TN 37831, USA
}%

\author{N.T. Howard}
\author{E.S. Marmar}

\affiliation{ 
Massachusetts Institute of Technology, Cambridge, MA, 02139, USA
}%

\author{S. Mordijck}
\affiliation{ 
Department of Physics, College of William \& Mary, Williamsburg, VA, USA
}%

\date{\today}%




\begin{abstract}
We present Aurora, an open-source package for particle transport, neutrals and radiation modeling in magnetic confinement fusion plasmas. Aurora's modern multi-language interface enables simulations of 1.5D impurity transport within high-performance computing frameworks, particularly for the inference of particle transport coefficients. A user-friendly Python library allows simple interaction with atomic rates from the Atomic Data and Atomic Structure database as well as other sources. This enables a range of radiation predictions, both for power balance and spectroscopic analysis. We discuss here the \emph{superstaging} approximation for complex ions, as a way to group charge states and reduce computational cost, demonstrating its wide applicability within the Aurora forward model and beyond. Aurora also facilitates neutral particle analysis, both from experimental spectroscopic data and other simulation codes. Leveraging Aurora's capabilities to interface SOLPS-ITER results, we demonstrate that charge exchange is unlikely to affect the total radiated power from the ITER core during high performance operation. Finally, we describe the \texttt{ImpRad} module in the OMFIT framework, developed to enable experimental analysis and transport inferences on multiple devices using Aurora.
\end{abstract}

\maketitle

\section{Introduction} \label{sec:intro}
As we approach the era of burning plasmas in fusion devices, advancing predictive capabilities and validating models of particle transport, impurity radiation and neutral particles has become increasingly important. The scaling of fusion power with fuel density squared makes an obvious case for the need to understand density peaking. Effective operation also requires the development of models that may help in controlling impurity accumulation near axis and actively mitigating divertor heat fluxes via edge puffing. Appreciating the important interactions of neutrals particles, both in terms of fueling and as partners in charge exchange reactions, is also fundamental for core-edge integration. One common thread through these subjects is the need to simultaneously consider the roles of both transport and atomic physics. In this paper, we present the Aurora package, which aims to address some of these needs of the fusion community by providing a modern toolbox for impurity transport modeling, radiation predictions and spectroscopic analysis. 

Aurora is well integrated with Python tools originally developed within the One Modeling Framework for Integrated Tasks~\cite{Meneghini2015IntegratedOMFIT} (OMFIT) and recently released for independent installation. This allows it to leverage a wide community effort to generalize common analysis routines and access data from multiple devices. For example, reading of files in standardized formats, e.g. from EFIT~\cite{Lao1985ReconstructionTokamaks} or GACODE routines, is easily accomplished, permitting non-trivial manipulations of experimental data without need for detailed expertise. Integration of Aurora within integrated modeling frameworks is made particularly straightforward by its simple installation procedures, multi-language interfaces and device-agnostic implementation. Aurora is an entirely open source project, hosted on \emph{GitHub} and welcoming contributions from users with a range of interests. Documentation is available at \url{https://aurora-fusion.readthedocs.io}.
\\ \\
In this paper, we aim to describe the current status of the project and provide a clear reference for some key features. Section~\ref{sec:atomic} gives an overview of capabilities related to atomic data and radiation predictions, particularly making use of the Atomic Data and Analysis Structure (ADAS) database~\cite{Summers2006IonizationElements}. In Section~\ref{sec:forward_model} we outline Aurora's 1.5D simulation capabilities for ion transport in magnetically confined plasmas. Section~\ref{sec:superstaging} explains the concept of \emph{superstaging} and its use in Aurora as a way to reduce the numerical complexity of heavy ions and to test approximations for edge codes. In section~\ref{sec:cxr} we discuss how processing of SOLPS-ITER results with Aurora indicates that charge exchange (CX) with edge neutrals should not significantly affect the total radiated power, $P_{\text{rad}}$, within the last closed flux surface (LCFS) in ITER. The subjects of superstaging and CX offer two examples of research activities where the modern interface and flexibility of Aurora has facilitated new insights, advancing both technical capabilities and physical interpretation for predictions relevant to future devices. In Section~\ref{sec:summary} we conclude by describing opportunities for experimental analysis using the new \emph{ImpRad} module in OMFIT, facilitating inferences of impurity transport on multiple devices based on Aurora.

\section{Atomic Data and Spectral Predictions} \label{sec:atomic}
Aurora relies on ADAS data for much of its functionality. While it may be convenient to combine usage of ADAS Python routines with Aurora, it is not necessary to run (or even have access to) an ADAS framework. Aurora can work with atomic data from the OPEN-ADAS website (\url{www.open-adas.ac.uk}), fetching and interpreting files automatically through the internet. Most of the Generalized Collisional-Radiative (GCR) formats distributed by ADAS can be processed. While users may indicate specific ADAS files that they wish to use, a set of defaults are conveniently listed within Aurora for most ions of interest in fusion. This offers the side benefit of lowering the ``entry barrier'' for researchers who are not familiar with ADAS nomenclature and who may be unsure about which files offer the highest data quality. 
Cooling coefficients, both resolved for each ion stage and weighted by fractional abundances of each charge state at ionization equilibrium, can easily be loaded and processed for use in integrated modeling. For applications in spectroscopy, the \emph{aurora.read\_adf15} function allows parsing and interpolation of Photon Emissivity Coefficients (PECs) in the ADAS ADF15 format, separating components driven by ionization, excitation or recombination processes, as well as metastable-resolved parts whenever these are available in the chosen files. 
Aurora also offers initial support to work with the ColRadPy~\cite{Johnson2019ColRadPy:Solver} package, which makes use of ADAS resolved ion data collections in the ADF04 format for GCR model predictions. Specifically for the analysis of neutral H-isotope particles, some atomic data from the formulae by Janev \& Smith~\cite{Janev1993CrossIons} or Janev, Reiter \& Samm~\cite{Janev2003CollisionPlasmas} are also available. 

All the tools discussed in this section are applicable to Aurora simulations of particle transport (Section~\ref{sec:forward_model}) as well as other codes and analysis frameworks. Indeed, Aurora has been used for experimental spectroscopic signal analysis~\cite{Sciortino2020InferenceSelection, Sciortino2021ParticleLines} and to post-process data from neutral codes such as
KN1D~\cite{LaBombard2001KN1DPlasma}, FIDASIM~\cite{FIDASIMpaper, FIDASIMcode} and EIRENE~\cite{Reiter2005TheCodes}. Aurora does not include capabilities to work with molecules at the time of writing, but the subject will be explored in future work.

\section{1.5D Modeling of Particle Transport} \label{sec:forward_model}

Simulation of impurity transport is extremely useful for both the interpretation of experimental measurements as well as the assessment of radiative scenarios that reduce heat fluxes going into the boundary layer. A number of 1D/1.5D codes have been developed over the years for these purposes, including STRAHL~\cite{Behringer1987DescriptionSTRAHL, Dux2006STRAHLManual}, SANCO~\cite{Lauro-TaroniL.etal1994ImpurityJET}, MIST~\cite{Hulse1983NumericalPlasmas} and ITC~\cite{Parisot2008ExperimentalPlasma}. 
These codes were not designed to operate on high performance computing clusters, where one may want to keep data in memory rather than writing to disk. Since the task of writing data files can account for the majority of the runtime when dealing with large arrays, avoiding it can give a significant speed improvement, sometimes by a factor in excess of 10. Gaining such advantage was one of the original motivations for the development of Aurora as a tool to rapidly iterate over some input parameters in inferences of experimental impurity transport coefficients on Alcator C-Mod~\cite{Sciortino2020InferenceSelection} and DIII-D~\cite{Odstrcil2020DependenceTokamak}. In this context, Aurora's simulation capabilities are referred to as a ``forward model'', differentiating the task of obtaining numerical results from chosen inputs via a pre-defined model and the problem of iterating over forward model inputs until matching some termination conditions (``inverse problem''). 

Aurora's high-level Python interface provides simple operation and interpretability of results, for example by permitting plotting of atomic rates or the quantification of particle conservation. The high level interface does not come at the cost of reduced speed, since in practice the coupled transport equations for all ion charge states are solved in Fortran (via \emph{f2py}~\cite{Peterson2009F2PY:Programs}). Julia~\cite{bezanson2017julia} routines have also been recently added. 

\begin{figure}[!t]
	\centering
	\includegraphics[width=0.5\textwidth]{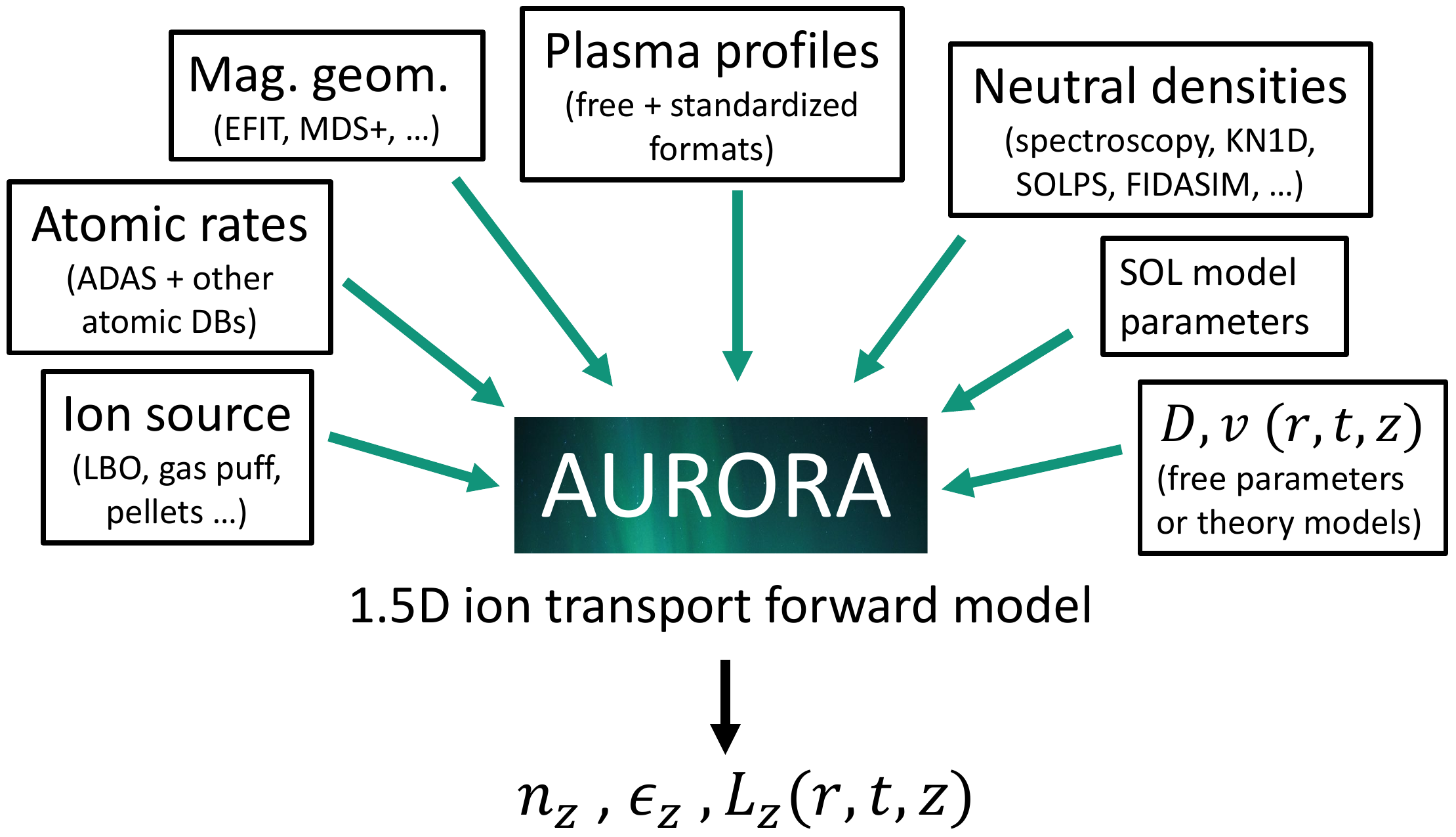}
	\caption{Overview of Aurora's 1.5D ion transport forward model. All inputs can be obtained either from experimental data or modeling, except for atomic rates (normally from ADAS) and transport coefficients (either from a theoretical transport model or considered ``free parameters''). Aurora can calculate charge state densities ($n_z$), emissivities ($\epsilon_z$) and radiated power components ($L_z$).}
	\label{fig:AURORA_diagram}
\end{figure}

Aurora's forward model couples transport and atomic processes by solving continuity equations for each charge state over time and a 1D radial coordinate. Fig.~\ref{fig:AURORA_diagram} illustrates the main categories of inputs and outputs of the code, including possible sources for each input (either from experimental measurements or modeling). To enable investigation of non-circular plasma shapes, the forward model spatial coordinate is taken to be $r_{\text{vol}} = \sqrt{V/(2\pi^2 R_0)}$, where $V$ is flux surface volume and $R_0$ is the major radius at the magnetic axis. Note that this coordinate is not strictly ``radial'', since it encapsulates information about 3D flux surface volumes, and may also be applied in the context of stellarators~\cite{Geiger2019ObservationIron}. Since this spatial coordinate is numerically 1D but encapsulates information about higher-dimensional geometry, Aurora's forward model may be referred to as being 1.5D. The $r_{\text{vol}}$ coordinate is artificially extended into the Scrape-Off Layer (SOL) by using the major radii at the high- and low-field sides. While the resulting forward model is not appropriate for detailed considerations on edge transport, this extension into the SOL permits simple considerations on wall and divertor recycling using a small set of input parameters.

Aurora leverages the wide range of modeling capabilities originally developed within OMFIT, for example to process magnetic equilibria from EFIT output files and read standard file formats. Aurora's integration with OMFIT routines also ensures that it lives within a software framework that is continuously expanded and adapted to challenges in fusion research over time.

The objective of Aurora's forward model is to solve a set of coupled continuity equations for an ion ``I'' and all its charge states ``Z'', which in cylindrical geometry can be written as 
\begin{equation} \label{eq:nz_cont}
\frac{\partial n_{I,Z}}{\partial t} = - \frac{1}{r} \frac{\partial}{\partial r} \left( r \Gamma_{I,Z} \right)+Q_{I,Z}.
\end{equation} 
The radial (cross-field) particle flux $\Gamma_{I,Z}$ is separated into diffusive and convective terms with the standard ansatz
\begin{equation} \label{eq:rad_flux} 
\Gamma_{I,Z}=-D \frac{\partial n_{I,Z}}{\partial r}+v \ n_{I,Z},
\end{equation}
where $D$ is the diffusion coefficient and $v$ is the radial convective velocity. We remark that $D$ and $v$ are \emph{inputs} to Aurora, not outputs. The coupling between charge states in Eq.~\ref{eq:nz_cont} occurs through the atomic sources/sinks term, $Q_{I,Z}$, which can be expanded into its components as
\begin{equation} \label{eq:atomic_source}
    \begin{split}
        Q_{I,Z} & = - (n_e S_{I,Z} + n_e \alpha_{I,Z} + n_H \alpha_{I,Z}^{cx}) \  n_{I,Z} \\ 
         &  \quad  + n_e S_{I,Z-1} \ n_{I,Z-1} \\ 
         &  \quad  + (n_e \alpha_{I,Z+1} + n_H \alpha_{I,Z+1}^{cx}) \ n_{I,Z+1}.
\end{split}
\end{equation}
The right hand side of Eq.~\ref{eq:atomic_source} is made of three terms multiplying adjacent charge states. Total ionization rates, obtained from the ADAS SCD data, are labelled as $S$. Radiative and dielectronic recombination rates are grouped into a rate labelled $\alpha$, obtained from ADAS ACD data. Finally, $\alpha^{\text{cx}}$ stands for the contribution from charge exchange of each impurity charge state with the atomic density of hydrogen-isotopes, $n_H$. Electron density is here labelled as $n_e$.

Theoretical transport models often make predictions at the low-field side (LFS) on a midplane radial coordinate, which we label as $r_m$. These results can be transformed to flux surface averaged (FSA) values of $D$ and $v$ for the Aurora forward model using~\cite{Angioni2014TungstenModelling}
\begin{equation} \label{eq:corrections}
D = D_{LFS} \frac{n_{LFS}}{\langle n\rangle} \left(\frac{\partial r_{\text{vol}}}{\partial r_m}\right)^2,
\end{equation}
\begin{equation*}
v = v_{LFS} \frac{n_{LFS}}{\langle n\rangle} \frac{\partial r_{\text{vol}}}{\partial r_m} + D \frac{\partial }{\partial r_m} \left( \ln \frac{n_{LFS}}{\langle n\rangle}\right)
\end{equation*}
where factors of $n_{LFS}/\langle n\rangle$ account for any poloidal density asymmetries. Centrifugal asymmetries can be particularly significant for the transport of heavy ions and may be estimated with Aurora given radial profiles of $n_e$, $T_e$ (electron temperature), $T_i$ (bulk ion temperature), $Z_{\text{eff}}$ (effective ion charge) and $\omega$ (toroidal rotation frequency)~\cite{Odstrcil2018TheUpgrade}.

Currently, Aurora uses by default a first-order, vertex-centered, finite-volume scheme recently developed by Linder~\cite{Linder2020Self-consistentUpgrade} which offers particle conservation and better numerical stability with respect to finite-difference schemes in the presence of large convection, as in pedestal regions of tokamaks. This scheme uses adaptive upwinding for the spatial discretization of the advective terms based on local evaluations at grid point ``i'' of the parameter
\begin{equation}
    K_{i}=\max \left(0,1-2 /\left|\mu_{i}\right|\right) \cdot \operatorname{sgn}\left(\mu_{i}\right),
\end{equation}
where $\mu_{i}=\left|v\left(r_{i}\right)\right| \Delta r_{i} / D\left(r_{i}\right)$ is the local P\'eclet number and $\Delta r_{i}$ is the local radial grid spacing, which can be flexibly set to be smaller in the pedestal. In the limit of $\mu_{i}\rightarrow 0$ ($K_i=0$, diffusion-dominated case) a purely central scheme is used, while for $\mu_{i}\rightarrow \infty$ ($K_i=1$, advection-dominated case) pure upwinding is adopted. It is recommended that users apply a highly non-uniform radial grid, maintaining $|\mu_{i}| \leq 2$ at all locations and thus avoiding numerical damping. Suggestions on how to do so are provided in the Aurora documentation.

The temporal discretization is performed using the $\theta$-method, equally weighting contributions from densities at the previous time step and the new time step (i.e. $\theta=1/2$). Ionization and recombination rates are set to act on each charge state using the Lackner method~\cite{Lackner1982AnRates}, which considers ionization and recombination in two half steps. In the first half step, the ionization term is computed using the density at the new time step (implicitly) and recombination with the density at the previous time step (explicitly); in the second half step, ionization is computed at the previous time step and recombination at the new time step. This method is unconditionally 
stable and allows one to use the fast Thomas algorithm to solve the resulting tri-diagonal system of equations for the charge states. The computational cost of this scheme scales linearly with the number of evolved charge states. Readers are referred to Refs.~\onlinecite{Linder2020Self-consistentUpgrade} and~\onlinecite{Lackner1982AnRates} for details on the numerical scheme. Additionally, as proposed by Linder~\cite{Linder2020Self-consistentUpgrade}, Aurora can evolve the neutral stage in the same way as ionized stages, even though the specification of transport coefficients for neutrals is not trivial. Alternatively, users can choose not to evolve neutrals, in which case recombination for the first ionized stage is artificially set to zero to prevent particle losses. The 1st order finite differences scheme described in the STRAHL manual~\cite{Dux2006STRAHLManual} has also been implemented in Aurora to benchmark its routines against a more established code. Such benchmarks have proven to be perfectly satisfied when using identical modeling choices to those available in STRAHL.

In Aurora, $D$ and $v$ can be provided as a function of time, space and charge state (i.e. Z value). Typically, it is hard to determine what Z dependence should be assigned to transport coefficients, but in some cases this capability can enable validation efforts, e.g. to test the predictions of neoclassical transport in the pedestal region~\cite{Sciortino2020InferenceSelection}. It is also possible to provide transport coefficients for each charge state as obtained via a neoclassical and/or turbulent transport code.

In concluding this section, we summarize and highlight key advantages of the Aurora 1.5D forward model with respect to pre-existing codes:
\begin{itemize}
    \setlength\itemsep{0.0em}
    \item In iterative frameworks, Aurora can be more than $10\times$ faster. While the specific run time depends on user-specified spatio-temporal grids and number of ion stages, generally a simulation over 100~\si{ms} can take as little as a few \si{ms} of CPU time.
    \item The current forward model algorithm gives better particle conservation at large $v/D$ with respect to finite-difference schemes, even with half as many grid points.
    \item Interfacing with other modeling frameworks is made easier via dedicated modules and OMFIT tools.
    \item The \emph{superstaging} method (described in the next section) offers additional speed improvements and a simple test bed for atomic physics in edge transport codes.
\end{itemize}

\section{Superstaging} \label{sec:superstaging}
A significant challenge in investigations of impurity transport is the computational complexity of coupling atomic physics and transport for a large number of charge states. Particularly for edge codes, modeling of complex ions is often considered intractable. For simpler 1.5D modeling in the core, for example with Aurora's forward model, heavy ions like W make it hard to iterate over some free parameters with sufficient speed. However, with a number of high-Z machines already in existence and more under construction, such modeling is clearly of great interest. The \emph{superstage} approximation offers the means to reduce computational complexity with a simple and interpretable method.

The technique described here is analogous to one explored in past JET~\cite{Lauro-TaroniL.etal1994ImpurityJET,Foster2008OnPlasmas} and ITER~\cite{Bonnin2011Full-tungstenSOLPS} modeling. These works rely on the creation of ADAS atomic rates for superstages in a set of dedicated data files. This process has been significantly simplified in Aurora, where transformations of standard collisional-radiative coefficients are computed ``on the fly'', without need for additional files. The simplicity of these operations calls for an exposition of these methods and a reconsideration of research opportunities associated with superstaging. In this section, we focus on providing an intuitive understanding to make the technique more accessible to modelers. Analogous ideas may be applied to bundle metastable states, but we shall not discuss the subject here for simplicity.

In the superstaging approximation, multiple charge states are ``bundled'' together to reduce the number of species to evolve in a simulation. Suppose that we wish to group charge states $i_0$ to $i_1$ onto a superstage $\zeta$. We then have
\begin{equation} \label{eq:cs_sum}
    n^\zeta=\sum_{i=i_{0}}^{i_{1}} n^i
\end{equation}
where we have used lower case $n$ for the density of a general atomic species. Superscripts indicate the stage or superstage of interest, in slight departure from the more common notation used in section~\ref{sec:forward_model}. In what follows, electron density is simply labeled as $n_e$. The continuity equation for a charge state $i$ can be written as
\begin{equation} \label{eq:cont_cs_super}
    \frac{\partial n^i}{\partial t}=- \nabla \cdot \Gamma^i+Q^i.
\end{equation}
and our objective is to obtain an equation of identical form, but with indices $i$ replaced by $\zeta$ and each term appropriately adapted to represent superstages. Eventually, after solving for the superstage densities, we wish to split these back into individual charge states densities.

Let us assume that the transport coefficients appearing in the $\Gamma^i$ term in Eq.~\ref{eq:cont_cs_super} are independent of $Z$; we shall come back to this approximation later. Summing particle flux terms of multiple charge states in Eq.~\ref{eq:cont_cs_super} is then trivial, since $D$ and $v$ are the same for all charge states and densities linearly sum up as in Eq.~\ref{eq:cs_sum}. The same is true for the time derivatives of particle density on the left hand side of Eq.~\ref{eq:cont_cs_super}. Let us therefore focus our attention on the sources/sinks term, $Q^i$, which we write here for a charge state $i$ as
\begin{equation} \label{eq:atomic_source}
    \begin{aligned}
    Q^i = &- (n_e S^{i\rightarrow i+1} + n_e \alpha^{i\rightarrow i-1}) \  n^i \\
    &+ n_e S^{i-1\rightarrow i} \ n^{i-1} + n_e \alpha^{i+1 \rightarrow i} \ n^{i+1}.
    \end{aligned}
\end{equation}
Here we have adopted a notation that makes it clear which charge states are involved in each reaction, e.g. $S^{i\rightarrow i+1}$ is the process of ionization from charge state $i$ to $i+1$ and $\alpha^{i\rightarrow i-1}$ is recombination from charge state $i$ to $i-1$. Upon summation of the $Q^i$ term for each of the charge states $i_0$ to $i_1$, many terms cancel out. The sum over these charge states can be written as
\begin{equation} \label{eq:Q_i_superstage}
    \begin{aligned}
    Q^\zeta = &+ n_{e} \ S^{i_{0}-1 \rightarrow i_{0}} n^{i_{0}-1} \\
    &-n_{e} \alpha^{i_{0} \rightarrow i_{0}-1} n^{i_{0}}  \\
    &-n_{e} S^{i_{1} \rightarrow i_{1}+1} n^{i_{1}}  \\
    &+n_{e} \alpha^{i_{1}+1 \rightarrow i_{1}} n^{i_{1}+1}.
    \end{aligned}
\end{equation}

The purpose of superstaging is to avoid explicit modeling of individual charge states, which must instead be only related to the correspondent (directly modelled) superstages. \emph{In the superstage approximation, we impose quasi-static ionization equilibrium for the charge states $i_{0}$ to $i_{1}$ within a superstage}. This means that the fractional abundance of each charge state within the superstage is purely set by atomic physics, with no transport involved. Within a superstage, we therefore take
\begin{equation} \label{eq:superstages_frac_abund}
    \begin{aligned}
    \left.n^{i_{0}}\right|_{e q} &=\left.\left(\alpha^{i_{0}+1 \rightarrow i_{0}} / S^{i_{0} \rightarrow i_{0}+1}\right) n^{i_{0}+1}\right|_{e q} \\
    \left.n^{i_{0}+1}\right|_{e q} &=\left.\left(\alpha^{i_{0}+2 \rightarrow i_{0}+1} / S^{i_{0}+1 \rightarrow i_{0}+2}\right) n^{i_{0}+2}\right|_{e q} \\
    & \cdots \\
    \left.n^{i_{1}-1}\right|_{e q} &=\left.\left(\alpha^{i_{1} \rightarrow i_{1}-1} / S^{i_{1}-1 \rightarrow i_{i}}\right) n^{i_{1}}\right|_{e q}.
    \end{aligned}
\end{equation}
In other words, we wish to evolve $n^\zeta$ in space and time, i.e. $d n^\zeta/dt\neq 0$ in a transport code, and we take $Q^\zeta = 0$ when decomposing each superstage density into its charge state components. This obviously incurs an error, which however is small in many circumstances, as we will show later.

\begin{figure*}[!htb]
	\centering
	\includegraphics[width=0.8\textwidth]{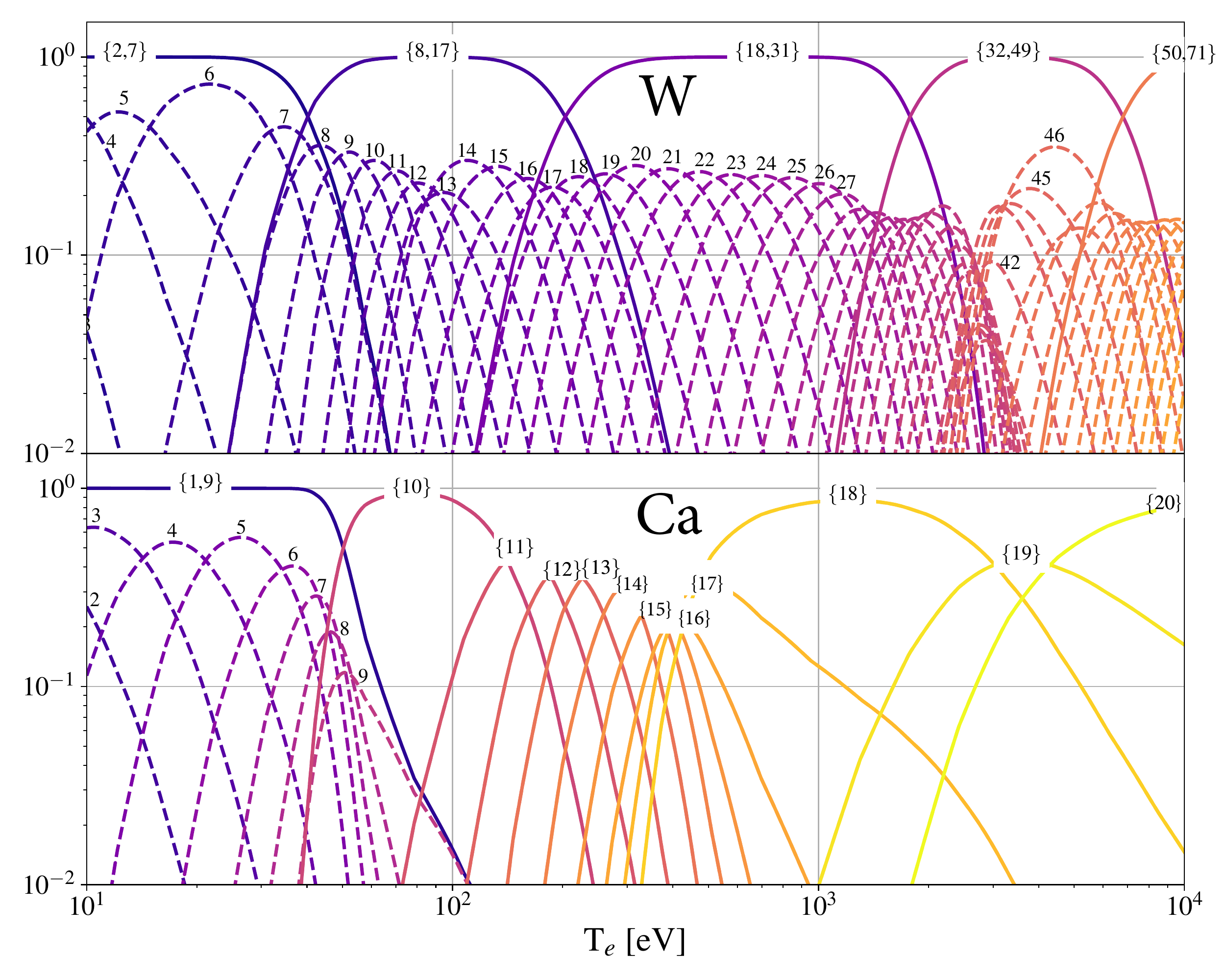}
	\caption{Full ionization equilibria of (top) W and (bottom) Ca over electron temperature, $T_e$, shown by dashed lines. Numbers within the plot identify selected ionization stages. Continuous  lines show one bundling scheme for each atomic species, in the case of W based on filled electronic shells, while for Ca only low-Z stages are grouped.}
	\label{fig:W_Ca_frac_abundances_superstaging}
\end{figure*}

In order to express Eq.~\ref{eq:Q_i_superstage} solely in terms of superstage properties, we recognize that both charge states $i_0$ and $i_1$ are absorbed into superstage $\zeta$, so by applying the changes $i_0\rightarrow \zeta$ and $i_1\rightarrow \zeta$ one can write
\begin{equation} \label{eq:superstage_rels}
    \begin{aligned}
    S^{i_{0}-1 \rightarrow i_{0}} n^{i_{0}-1} & \rightarrow S^{\zeta-1 \rightarrow \zeta} n^{\zeta-1} \\
    \alpha^{i_{0} \rightarrow i_{0}-1} n^{i_{0}} & \rightarrow \alpha^{\zeta \rightarrow \zeta-1} n^{\zeta} \\
    S^{i_{1} \rightarrow i_{1}+1} n^{i_{1}} & \rightarrow S^{\zeta \rightarrow \zeta+1} n^{\zeta} \\
    \alpha^{i_{1}+1 \rightarrow i_{1}} n^{i_{1}+1} & \rightarrow \alpha^{\zeta+1 \rightarrow \zeta} n^{\zeta+1}
    \end{aligned}
\end{equation}
which allow us to define effective superstage rates as
\begin{equation} \label{eq:superstage_alpha}
    \alpha^{\zeta \rightarrow \zeta-1} \equiv \alpha^{i_{0} \rightarrow i_{0}-1} \left(\frac{n^{i_{0}}}{n^{\zeta}} \right)_{eq},
\end{equation}
\begin{equation} \label{eq:superstage_S}
    S^{\zeta \rightarrow \zeta+1} \equiv S^{i_{1} \rightarrow i_{1}+1}\left(\frac{n^{i_{1}}}{n^{\zeta}}\right)_{eq}.
\end{equation}
The terms in brackets are the fractional abundances of the ``edges'' of the superstage $\zeta$, which we explicitly labelled as being computed from ionization equilibrium. Using the superstage rates of Eqs.~\ref{eq:superstage_alpha} and~\ref{eq:superstage_S}, one can then express the sources/sinks term for superstages as
\begin{equation} \label{eq:Q_i_superstage_2}
    \begin{aligned}
    Q^\zeta = &- (n_e S^{\zeta\rightarrow \zeta+1} + n_e \alpha^{\zeta\rightarrow \zeta-1}) \  n^\zeta \\
    &+ n_e S^{\zeta-1\rightarrow \zeta} \ n^{\zeta-1} + n_e \alpha^{\zeta+1 \rightarrow \zeta} \ n^{\zeta+1}.
    \end{aligned}
\end{equation}
which has the same form as Eq.~\ref{eq:atomic_source} and can be used for impurity transport simulations, including (but not exclusively) within the 1.5D Aurora forward model. 

$D$ and $v$ can either be set to be the same for all superstages, as commonly done in simulations with individual charge states, or users may indicate an arbitrary $Z$ dependence. This could, for example, be obtained from a neoclassical or turbulent transport code, in which case Aurora can compute an average of transport coefficients weighted by the fractional abundance of charge states within each superstage. This corresponds to taking 
\begin{equation}
    D^\zeta = \sum_{i_0}^{i_1} D^i \left(\frac{n^i}{n^{\zeta}}\right)_{eq}.
\end{equation}
An analogous weighing by fractional abundances at ionization equilibrium allows one to decompose superstages into individual charges after a simulation:
\begin{equation} \label{eq:unstaging}
    n^i = n^\zeta \left(\frac{n^i}{n^{\zeta}}\right)_{eq}.
\end{equation}
We remark that while $n^\zeta$ may be significantly out of ionization equilibrium, resulting from a simulation with finite transport, the right hand side term within brackets in Eq.~\ref{eq:unstaging} is taken purely from ionization equilibrium (no transport).

Once superstage densities have been unstaged, any calculation of radiation terms can proceed as usual using standard collisional-radiative coefficients, with no need for additional tabulation of atomic rates. Superstaging is therefore seen to only require charge state ionization and recombination rates. If charge exchange is considered, the ADAS CCD data are also needed for inclusion within the $\alpha$ rates with appropriate weighting by $n_n/n_e$, where $n_n$ is the background H-isotope atomic neutral density. 

The error incurred by the superstaging approximation is dependent on the intensity of plasma transport, since this technique effectively imposes ionization equilibrium within superstage partitions. For typical tokamak transport levels, there often exist multiple useful and safe partitioning strategies for medium- and high-Z impurities. It is however not possible to specify the ``best'' partition in a universal manner, since the optimal choice of superstages depends on the application of interest. The top panel of Fig.~\ref{fig:W_Ca_frac_abundances_superstaging} shows W fractional abundances of all charge states of W as a function of $T_e$, using dashed lines. Easily identifiable charge states have been indicated by numbers near the peak fractional abundance of each. A partition of charge states based on the valence electronic shells is also shown by continuous lines, corresponding to superstages $Z \in \{0\}$, $\{1,2\}$, $\{3,8\}$, $\{9,18\}$, $\{19,32\}$, $\{33,50\}$, $\{51,72\}$, $\{72,74\}$. This partitioning is physically motivated by atomic physics and analogous to the ``natural partition'' strategy defined in ADAS, whereby charge states are bundled based on variations of ionization potential ($I_z$) between charge states. Specifically, the natural partition is computed for each species by finding large deviations of the quantity $2 (I_{z+1}-I_z)/(I_{z+1}+I_z)$ from a running mean over charge states~\cite{Foster2008OnPlasmas}. To this end, Aurora can make use of ionization energies from NIST~\cite{NIST_ASD}, collected via ColRadPy~\cite{Johnson2019ColRadPy:Solver}. Alternatively, users may also use fractional abundances at ionization equilibrium to determine which charge states have a small range of existence in a given simulation and can therefore be safely bundled. Such ``clustering'' of superstages can be easily dealt with using the K-means algorithm.

However, often the optimal partitioning of charge states may be specific to the application of interest. For example, for edge simulations, charge states with charge greater than W20+ are unlikely to be useful and can always be bundled. Similarly, core simulations may bundle low- and high-lying charge states that are never reached at the temperatures of interest, only simulating charge states that produce observable effects. We note that the 0th charge state is always needed in forward modeling where ion sources are introduced in the neutral state. The bottom panel of Fig.~\ref{fig:W_Ca_frac_abundances_superstaging} shows a case where we bundle charge states $\{1,9\}$ and keep individual charge states with $Z>9$. This choice is favorable for core impurity transport studies where Ca is injected into tokamak plasmas via Laser Blow Off (LBO) since emission from charge states 1-9 is typically not measured and only higher charge states must be carefully resolved. In tests of the Aurora forward model with this partitioning strategy, we have found that a negligible error in impurity densities and radiation is incurred for any realistic transport levels. Recall that the computational cost of the Aurora forward model is linearly proportional to the number of modelled charge states, so superstaging as in Fig.~\ref{fig:W_Ca_frac_abundances_superstaging} gives approximately a factor of 2 speed improvement.

\begin{figure}[htp]

    \subfloat[]{%
      \includegraphics[clip,width=0.9\columnwidth]{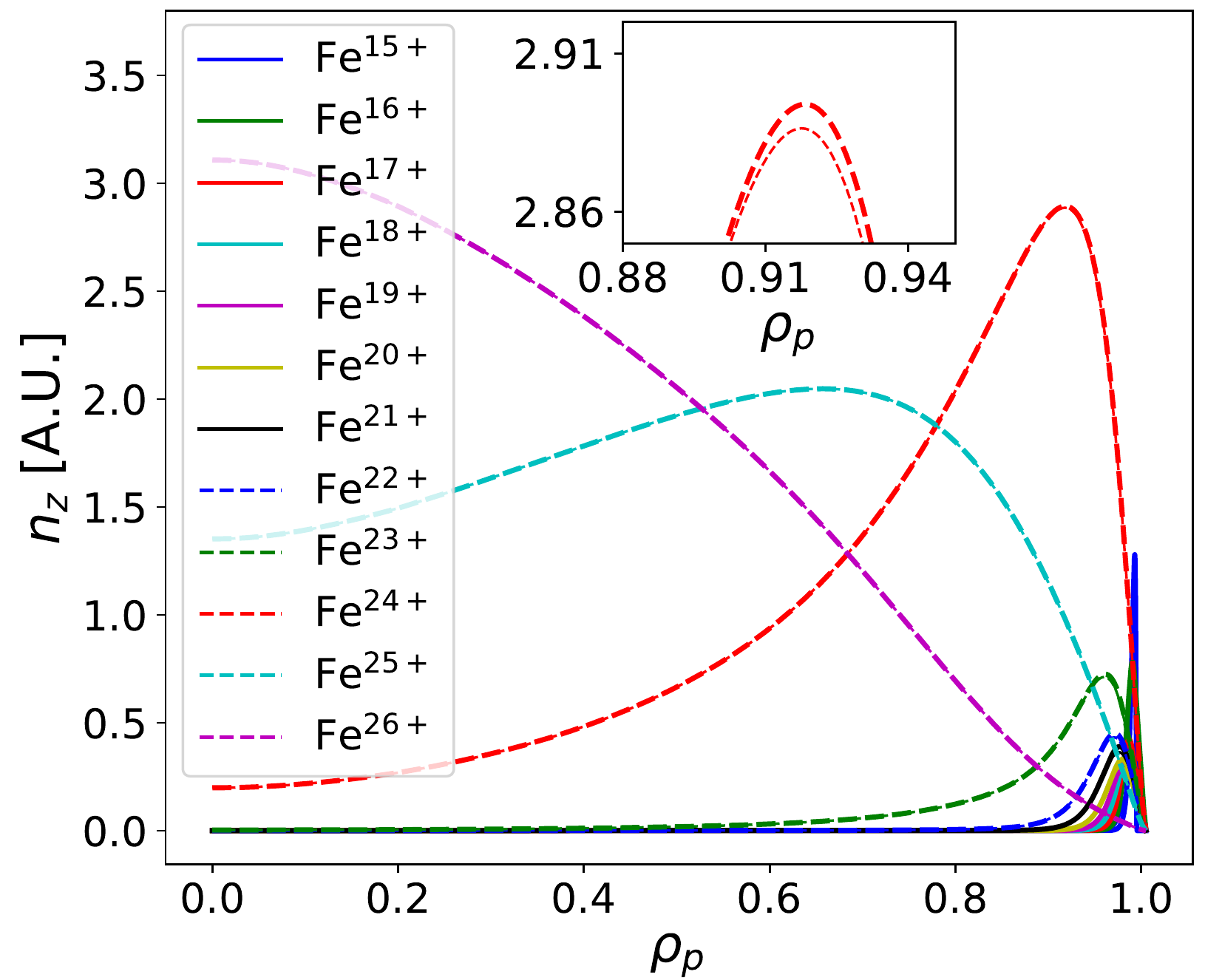}%
      \label{fig:Fe_sim_superstaging_ITER}
    }
    
    \subfloat[]{%
      \includegraphics[clip,width=0.9\columnwidth]{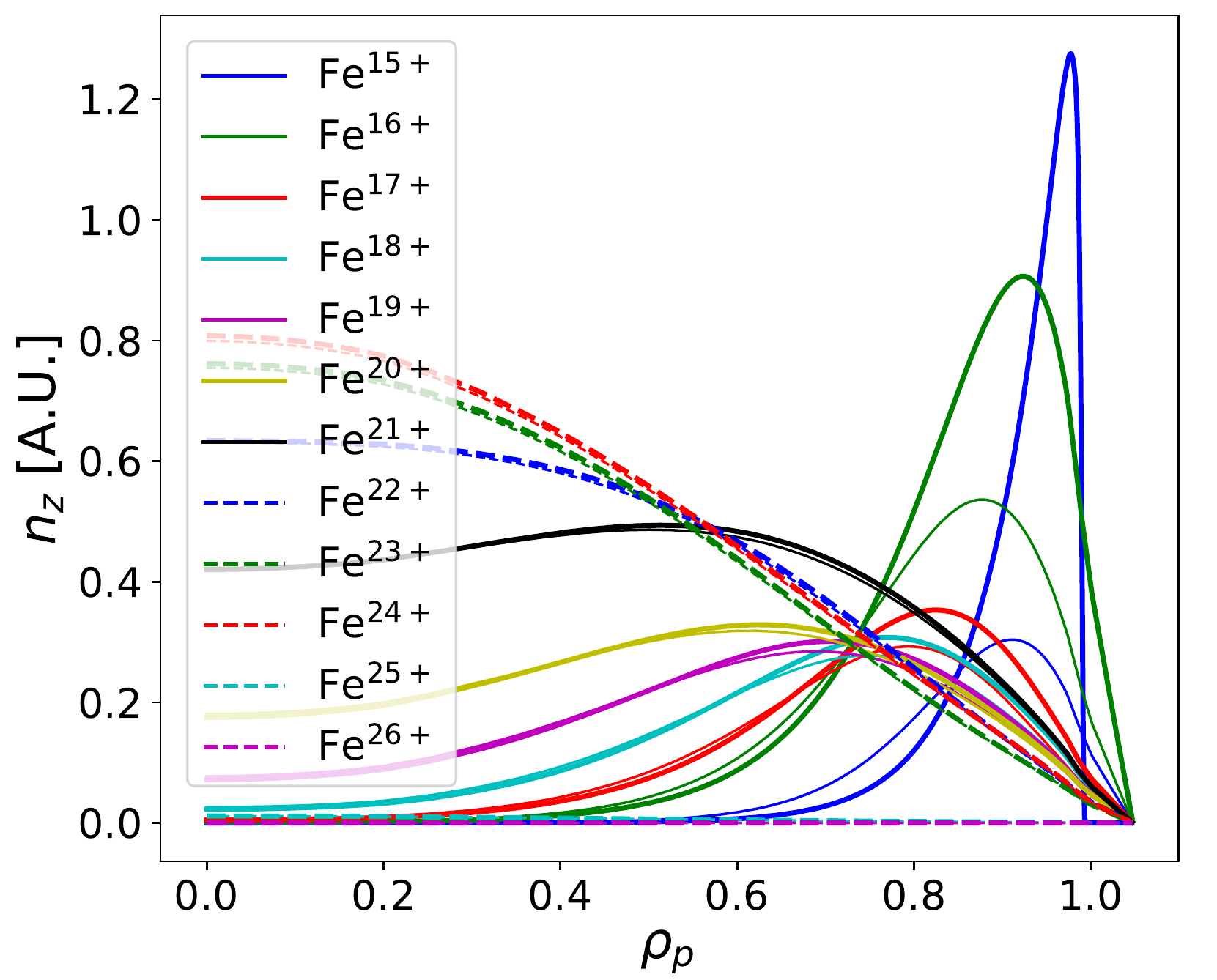}%
      \label{fig:Fe_sim_superstaging_CMOD}
    }
    
    \caption{Comparison of Aurora simulations for Fe transport in (a) the ITER baseline scenario ($T_e\approx 25$ \si{keV} and $n_e\approx 1.1\times 10^{20}$ \si{m^{-3}} on axis), and (b) an Alcator C-Mod discharge with $T_e\approx3.5$ \si{keV} and $n_e\approx 1.5\times 10^{20}$ \si{m^{-3}} on axis. Transport coefficients $D=10$~\si{m^2/s}, flat across the radius, and $v=0$~\si{m/s} were assumed for illustrative purposes. Thin lines show results from a simulation resolving all charge states, thick lines from bundling $Z<16$ states.}
    \label{fig:Fe_sim_superstaging}
\end{figure}

Fig.~\ref{fig:Fe_sim_superstaging_ITER} shows two Aurora simulations for Fe injection in the ITER baseline scenario, where peak electron temperatures and densities reach $T_e\approx 25$ \si{keV} and $n_e\approx 1.1\times 10^{20}$ \si{m^{-3}}. We compare the standard use of all charge states (thin lines) with a superstage partition that bundles Fe charge states with $Z<16$ (thick lines), analogously to the Ca strategy of Fig.~\ref{fig:W_Ca_frac_abundances_superstaging}. Almost no difference is observed in the charge state densities of interest; the inset shows a zoomed in view of the peak of Fe$^{24+}$ near the pedestal, where a small inaccuracy is seen. For this comparison, a flat diffusion coefficient with magnitude $D=10$~\si{m^2/s} was assumed, setting convection to zero for simplicity. We note that the value of $D=10$~\si{m^2/s} is relatively high and, since superstaging becomes less accurate at high transport levels, this can be seen as a successful test in challenging, yet reasonable, conditions for the application of this method. We remark however that the same partitioning strategy can be much less successful in other plasma scenarios. Fig.~\ref{fig:Fe_sim_superstaging_CMOD} shows the result of using the same Fe superstaging strategy for realistic Alcator C-Mod kinetic profiles with on-axis values of $T_e\approx3.5$~\si{keV} and $n_e\approx 1.5\times 10^{20}$~\si{m^{-3}}, much lower than in the ITER baseline. As in Fig.~\ref{fig:Fe_sim_superstaging_ITER}, thin lines correspond to the simulation resolving all charge states, while thicker lines show the result of bundling all $Z<16$ states (except the neutral stage, as always). While core profiles are recovered very well, significant discrepancies are found in the outer part of the plasma. Of course, even larger errors would be incurred if a smaller number of superstages were directly modelled, e.g. taking only every other Fe state between $Z=16$ and $Z=26$.

\begin{figure}[htp]

    \subfloat[]{%
      \includegraphics[clip,width=0.9\columnwidth]{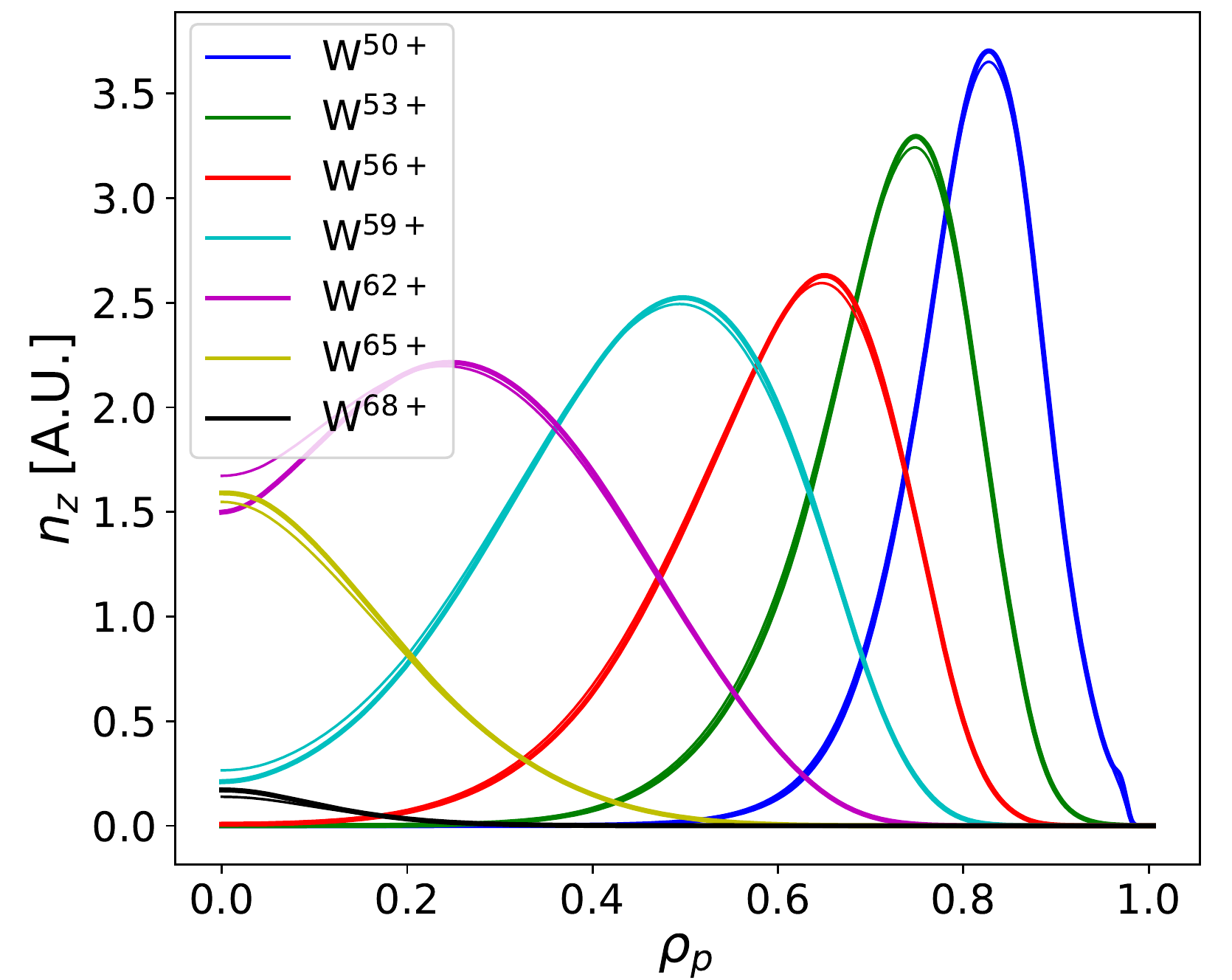}%
      \label{fig:W_sim_superstaging}
    }
    
    \subfloat[]{%
      \includegraphics[clip,width=0.9\columnwidth]{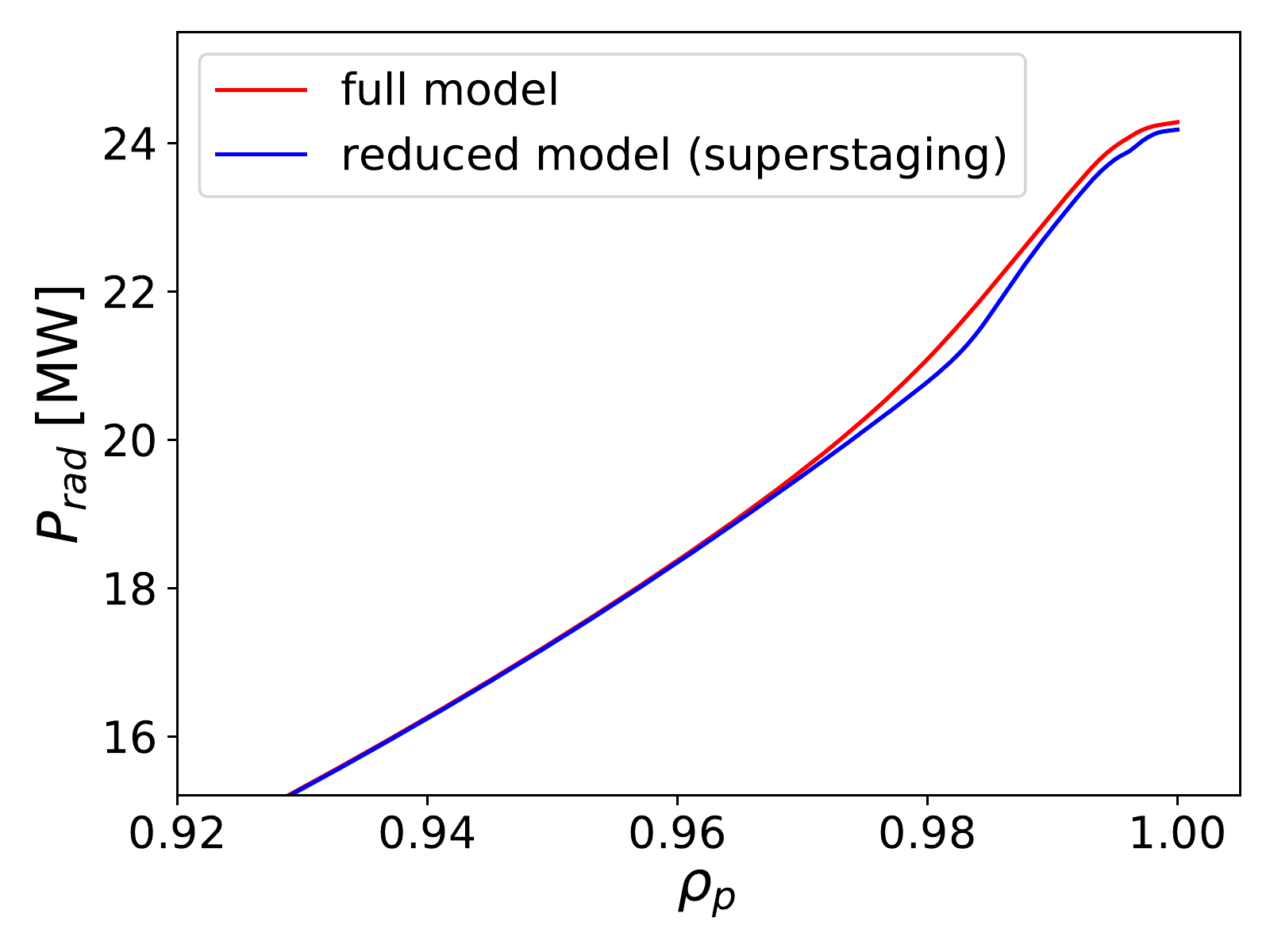}%
      \label{fig:W_sim_rad_superstaging}
    }
    
    \caption{Application of superstaging within an Aurora simulation of W in the ITER baseline scenario. In (a), thin lines show the result of simulating all charge states, while thicker lines show the result of superstaging. Here, only $Z=\{50,53,56,59,62,65,68\}$ were superstaged, using $D=10$~\si{m^2/s}, $v=0$~\si{m/s}. In (b), we show the difference between the two cases for cumulative radiated power, integrated from the magnetic axis outwards, in the pedestal region.}
\end{figure}

Many of the envisioned applications of superstaging are with high-Z ions. Fig.~\ref{fig:W_sim_superstaging} shows two Aurora simulations of W in the ITER baseline scenario. The displayed charge states were the only ones that were explicitly modelled, i.e. all others were bundled. Except for small inaccuracies near axis, the agreement between thin and thick lines is excellent, particularly considering the large reduction of computational cost of this superstaging partition. Fig.~\ref{fig:W_sim_rad_superstaging} shows the cumulative radiated power computed by Aurora as a function of plasma radius when setting the total W concentration to be $1.5\times 10^{-5}$ on axis~\cite{Putterich2019DeterminationFactors}. The radial range is focused on the pedestal region in order to better visualize the loss of accuracy due to superstaging, shown by the difference between the full model, simulating all charge states (red) and the reduced superstaged model (blue). Clearly, such simulation strategy could be very advantageous for a number of applications, including integrated modeling where impurity radiation can play a significant role. In these cases, modeling impurities as having charge state densities set by a constant fraction of the electron density can result in significant errors, whereas running Aurora with a small number of superstages, appropriately chosen for the scenario of interest, can be fast and effective.

\section{Charge Exchange in the ITER Pedestal} \label{sec:cxr}
Aurora's routines for the post-processing of SOLPS-ITER~\cite{Wiesen2015ThePackage,Bonnin2016PresentationModelling} results enable consideration of core-edge effects, for example related to the influence of edge neutrals in the pedestal region~\cite{Sciortino_NF_2021}. Recent work by Dux \emph{et al}~\cite{Dux2020InfluenceUpgrade} demonstrated the large impact of charge exchange on ionization balance in the AUG pedestal, making use of detailed Charge eXchange Recombination Spectroscopy (CXRS) measurements of Ne$^{8+}$ and Ne$^{10+}$ and a 1D Monte Carlo neutral code to model edge neutral penetration. An extrapolation to ITER based on these AUG observations suggested that CX could significantly affect the total radiated power emitted within the LCFS, $P_{\text{rad}}$, increasing core Ne radiation by up to a factor of 5. Here, we provide a new assessment of this subject based on EIRENE modeling of neutral penetration in ITER using the SOLPS-ITER code, which is a higher-fidelity model with respect to previous descriptions.
\begin{figure}[!b]
	\centering
	\includegraphics[width=0.4\textwidth]{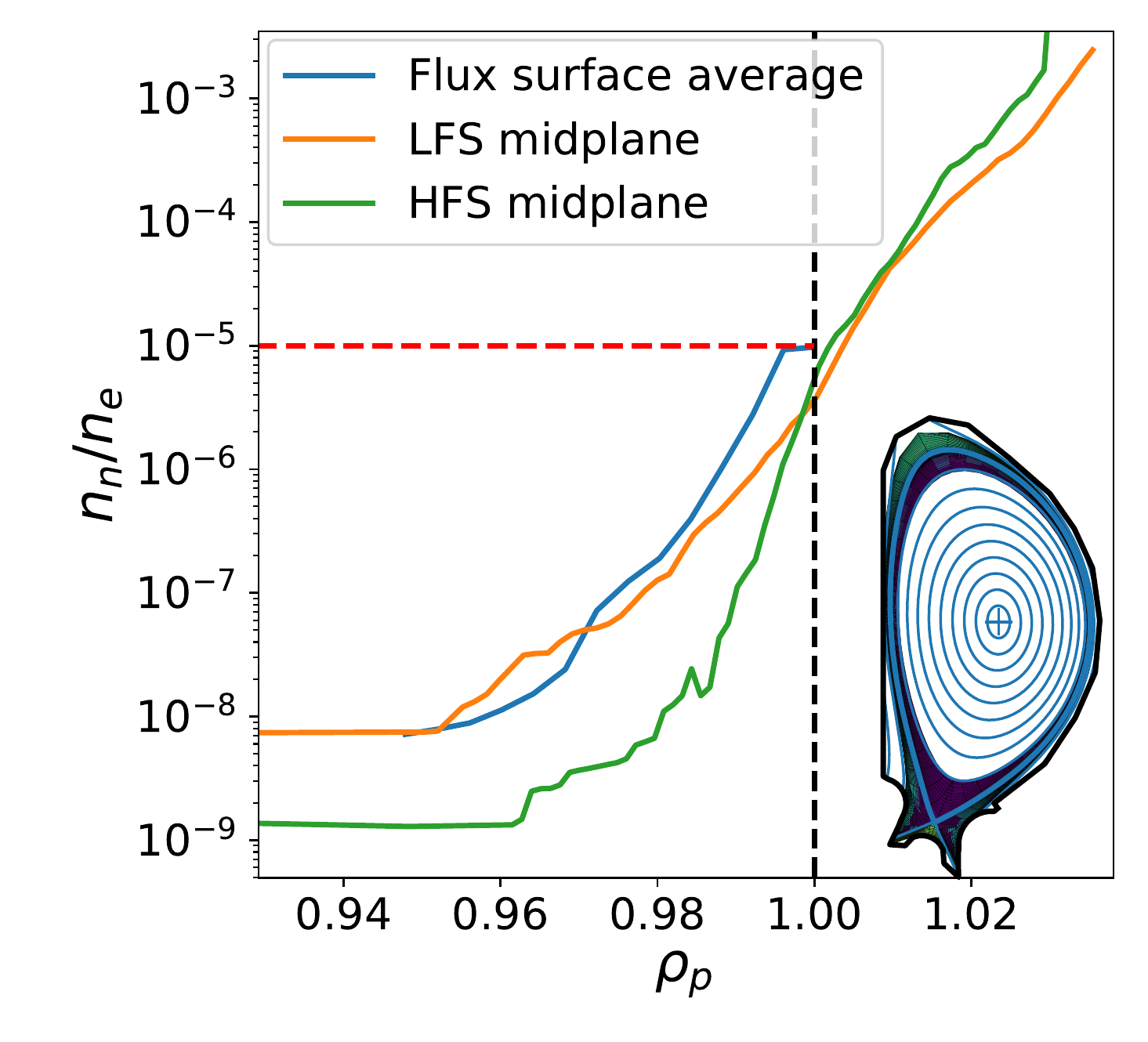}
	\caption{Radial profiles of atomic neutral D density, extracted from the SOLPS-ITER results displayed in the inset, showing interpolated slices at the low-field side (LFS) and high-field side (HFS) midplane, as well as a flux surface average. The dashed black line identifies the LCFS and the dashed red line shows the level of $n_n/n_e=10^{-5}$, below which CX does not significantly affect the local radiated power.}
	\label{fig:iter_solps_1d_nn_by_ne_radial_slices}
\end{figure}
The details of these SOLPS-ITER results, including realistic levels of Ne puffing for divertor heat flux mitigation, have been presented elsewhere as part of ITER divertor studies~\cite{Pitts2019PhysicsDivertor}. Here, we limit ourselves to a demonstration of post-processing and interpretation of the SOLPS-ITER data using Aurora. By providing the path to the SOLPS-ITER results directory, including an EFIT gEQDSK file, Aurora can load data on both the EIRENE and B2 grids with just a couple of lines of Python code. An additional call to the \emph{aurora.solps.get\_radial\_prof} function allows extraction of low- (LFS) and high-field side (HFS) radial profiles at the midplane, as well as flux surface averages. Fig.~\ref{fig:iter_solps_1d_nn_by_ne_radial_slices} shows these atomic D neutral density profiles, normalized by the corresponding $n_e$ radial profiles (LFS, HFS and FSA). We note that FSA neutral profiles are appropriate inputs to Aurora's 1.5D impurity transport model to examine the effect of neutrals on impurity ionization balance via CX. This is a physical modeling choice everywhere except very close to the LCFS (typically, for $\rho_p<0.99$) since the parallel motion of impurities on flux surfaces tends to be much faster than their ionization. As a result, the impact of neutrals on impurity ions is effectively averaged out also in the presence of strong poloidal neutral asymmetries, e.g. near divertor x-points or neutral beam injection~\cite{Dux2020InfluenceUpgrade}. Fig.~\ref{fig:iter_solps_1d_nn_by_ne_radial_slices} shows that $n_n/n_e$ is expected to be lower than $10^{-5}$, i.e. below the red dashed line. At the realistic Ne densities of the ITER baseline scenario, Aurora predicts the contribution to $P_{\text{rad}}$ directly attributable to CX recombination to be only approximately $0.2$ MW. The indirect contribution from the modification of the Ne ionization equilibrium due to CX cannot be self-consistently assessed without turning CX on and off in the SOLPS-ITER simulation itself, but at the low values of $n_n/n_e$ this effect is undoubtedly very small. These conclusions can be interpreted as resulting from the large width of the ITER SOL, which significantly reduces neutral penetration into the confined region. We note that while Alcator C-Mod's highest density shots did reach similar opacity to high-performance ITER projections~\cite{Mordijck2020CollisionalityPlasmas}, the C-Mod SOL was significantly thinner. As a result, C-Mod's absolute neutral densities in the pedestal were lower than in any other current tokamak, but still higher than in ITER, whose core fueling is bound to have different global character.

\section{Discussion and Conclusions} \label{sec:summary}
In this paper we have presented Aurora, a package for particle transport, neutrals and radiation modeling. Beyond its original purpose as a modern impurity transport forward model, Aurora has found applications as a simple interface to atomic data from ADAS and other databases. It has also been used to post-process results from other codes and integrate them within a framework suited for core-edge research in magnetically-confined plasmas. Aurora's forward model inherits from pre-existing numerical methods, particularly developed at JET and AUG, with important new capabilities enabled by a high-level Python interface with Fortran, as well as initial support for Julia. In Section~\ref{sec:superstaging}, we presented the superstaging approach used in Aurora. This could be adopted well beyond the Aurora forward model, particularly benefiting computationally-intensive edge codes, offering a fast interface to test the superstaging assumption in the presence of finite cross-field transport. Section~\ref{sec:cxr} discussed the impact of CX on the total radiated power within the LCFS in ITER, making use of post-processing tools of SOLPS-ITER results within the package. The extremely low values of $n_n/n_e$ found inside the LCFS make CX negligible to the ITER $P_{\text{rad}}$. 

Many of Aurora's features are leveraged in the OMFIT \emph{ImpRad} module, where they enable fast simulations of impurity transport, radiation predictions, development of synthetic diagnostics, and inferences of impurity transport coefficients. A Graphical User Interface (GUI) allows the creation of detailed Aurora namelists~\footnote{https://aurora-fusion.readthedocs.io/en/latest/params.html}, simple loading of magnetic geometries and kinetic profiles from other OMFIT modules, visualization and post-processing of results, as well as benchmarking with STRAHL for some test cases. Complex models with time-dependent transport coefficients, possibly arising from sawteeth or Edge Localized Modes (ELMs), can be easily constructed via the GUI. A set of workflows for the inference of impurity transport coefficients from experimental data has also been implemented, permitting nonlinear optimization with LMFIT~\cite{lmfit}, Markov Chain Monte Carlo with the emcee package~\cite{Foreman-Mackey2013EmceeHammer} and nested sampling with MultiNest~\cite{Feroz2008MultimodalAnalyses}. These iterative frameworks make use of a set of synthetic diagnostics developed for a range of measurements, including from Extreme Ultra-Violet (EUV) spectroscopy, Soft X-Ray (SXR) arrays, Charge Exchange Recombination (CER), Visible Bremsstrahlung (VB), and bolometry. Inferences of particle transport can be submitted to remote clusters to access high-performance computing resources, where Aurora simulations can be run in parallel. The \emph{ImpRad} module has been generalized to minimize device-specific tasks and allow utilization of the framework on multiple devices.

We conclude by highlighting the open-source and collaborative nature of the Aurora project. In this paper, we have described and demonstrated some of its capabilities, showing how it builds on the success of previous modeling tools. This has led to new opportunities in fusion research, many of them naturally related to core-edge integration given the range of applicability of the current code base.

\begin{acknowledgments}
The authors wish to thank all OMFIT contributors for their team work towards developing shared software tools. FS would also like to thank L. Lauro-Taroni and A. Foster for helpful conversations on superstaging, R. Dux for making the STRAHL code and its manual available to us, and D. Sta\'nczak for facilitating online distribution of Aurora.

This material is based upon work supported by the Department of Energy under Award Numbers DE-SC0014264, DE-FC02-04ER54698,  DE-AC05-00OR22725 and DE-SC0007880. This work has been carried out within the framework of the EUROfusion Consortium and has received funding from the Euratom research and training program 2014-2018 and 2019-2020 under grant agreement No 633053. The views and opinions expressed herein do not necessarily reflect those of the European Commission.

\end{acknowledgments}

\section*{Data Availability Statement}
Code and documentation for the Aurora package are available at \url{https://aurora-fusion.readthedocs.io}.

\bibliography{references,references_2}

\end{document}